\documentclass[preprint]{aastex}

\usepackage{mathtext,bbm,amsmath,amsfonts,amssymb,indentfirst,syntonly,graphicx}
\usepackage[english]{babel}
\usepackage{slashbox}
\usepackage{calc}
\usepackage{tikz}
\usepackage{amsmath,amssymb,amsthm}
\usepackage{latexsym,graphicx,bbm}

\newcommand{\beq}{\begin{eqnarray}}
\newcommand{\eeq}{\end{eqnarray}}

\newcommand{\ba}{\left( \begin{array}}
\newcommand{\ea}{\end{array} \right)}

\shorttitle{Lorentz factor in the magnetic jet model}
\shortauthors{Chang et al.}

\begin{document}

\title{Constraining the Bulk Lorentz Factor of GRB Outflow\\in the Magnetic-dominated Jet Model}

\author{Zhe Chang\altaffilmark{1,2,*}, Hai-Nan Lin\altaffilmark{1, \dag}, Yunguo Jiang\altaffilmark{1,2,\ddag}}
\affil{\altaffilmark{1}Institute of High Energy Physics\\Chinese Academy of Sciences, 100049 Beijing, China}
\affil{\altaffilmark{2}Theoretical Physics Center for Science Facilities\\Chinese Academy of Sciences, 100049 Beijing, China}

\altaffiltext{*}{changz@ihep.ac.cn}
\altaffiltext{\dag}{linhn@ihep.ac.cn}
\altaffiltext{\ddag}{jiangyg@ihep.ac.cn}

\begin{abstract}
Recent observations by the {\it Fermi}-LAT showed that there are delayed arrivals of GeV photons relative to the onset of MeV photons in some GRBs. In order to avoid a large optical depth, the minimal value of the Lorentz factor has been estimated to be higher than 1000 in some brightest bursts. In this paper, we present a detailed calculation of the time delay between the MeV and GeV photons in the framework of the magnetic-dominated jet model. We find that the time delay strongly depends on the saturated bulk Lorentz factor of the jet. Inspired by this fact, we use this model to calculate the Lorentz factors of the four brightest {\it Fermi} bursts. The results indicate that the Lorentz factors are much smaller than that obtained from the ``single-zone" scenario. The short burst GRB 090510 has a minimal Lorentz factor 385, while the three long bursts GRB 080916c, GRB090902b and GRB 090926 have almost the same Lorentz factors, with an average value near 260.  Another interesting result is that, for long bursts, GeV photons are emitted after the bulk Lorentz factor saturates. For the short GRB, however, MeV and GeV photons are emitted at the same phase, i.e., either in the expansion phase or in the coasting phase.
\end{abstract}


\section{Introduction \label{sec:introduction}}

It is well known that photons with energy higher than $m_ec^2\approx0.511$ MeV in a local jet frame may annihilate into electron-positron pairs. The large optical depth of $\gamma\gamma$ annihilation, as well as the Compton scattering, will restrain these photons from escaping the jet. However, the observed GRB spectra often peak in the MeV range and sometimes extend to the GeV range. This is the so-called ``Compactness problem". Since the optical depth is proportional to the inverse of the bulk Lorentz factor ($\Gamma$) of the jet, the ``compactness problem" can be solved if we assume that the GRB outflow is moving with a large $\Gamma$ \citep{rees1966,piran1999}. The requirement of the thin optical depth for the observed high energy photons sets a lower limit on $\Gamma$ \citep{Lithwick2001}. The Burst and Transient Source Experiment (BATSE) data indicate that the Band-like spectra of most bursts have a thermal component at the prompt phase \citep{Ryde2004}, and the measurement of the temperature allows us to constrain both $\Gamma$ and the initial size  $r_0$ of the flow \citep{Pe'er:2007xm}.

The investigation of GRBs has entered into a new epoch since the launch of the {\it Fermi} satellite in June 2008. The {\it Fermi} LAT instrument has observed several GRBs with photon energy as high as tens GeV.  Within the framework of the simplified ``single zone" model, the GeV photons set a very large lower limit on $\Gamma$ \citep{Lithwick2001,Soderberg:2002yr,Razzaque2004,Granot2008,Abdo2009a,Abdo2009b,Abdo2009c,Ackermann2010,Ackermann2011,FermiLAT:2012ab,Ghisellini2010}. For example, \citet{Abdo2009a} showed that the minimal Lorentz factor of GRB 080916c outflow  was $\Gamma_{\rm min}\approx900$.  \citet{Ackermann2010} analyzed the spectra of GRB 090510 and showed that $\Gamma_{\rm min}\approx1200$. \citet{Ghisellini2010} estimated the decelerating time of GeV emissions and obtained the Lorentz factor of GRB 090510 as large as 2000. However, an efficient physical mechanism to boost the outflow to such a large $\Gamma$ is still unclear.

An interesting feature of the $Fermi$ observations is that GeV photons often arrived seconds later than  MeV photons \citep{Abdo2009a,Abdo2009b,Abdo2009c,Ackermann2010,Ackermann2011}. The explanation of this phenomenon should include both the intrinsic emission mechanism and the traveling process \citep{Chang:2012gq}. In some quantum gravity theories, photons can interact with the quantum fluctuation of the space-time, so high energy photons travel slower than low energy ones. Although this effect is very small, it can cause a detectable time difference after photons travel a cosmological distance \citep{Gambini:1998it,Ellis2008PLB,Ellis2011IJMPA}. Such Lorentz invariance violation (LIV) effects lead to a natural time delay between GeV and MeV photons \citep{Schaefer:1998zg,Abdo2009c,Boggs:2003kxa,Nemiroff:2011fk}. As it was shown by \citet{Abdo2009c} and \citet{Chang:2012gq}, the LIV effect is very small, we will neglect this effect in the following.

Without considering the LIV effects, the delayed arrival of GeV photons can also be explained by several GRB models \citep{Duran2011,Bosnjak:2011pt}. \citet{Duran2011} assumed that photons are emitted by electrons via synchrotron radiation, it takes more time for electrons to be accelerated to a large Lorentz factor in order to radiate GeV photons.  \citet{Bosnjak:2011pt} used the magnetic jet model, which was initially introduced by \citet{Drenkhahn2002} and \citet{Drenkhahn2002b}, to account for this phenomenon. According to this model, the optical depth is larger for high energy photons than that for low energy photons. GeV photons can only escape at a larger radius where the optical depth is below unity.

In the magnetic reconnection model, the Band-type spectra can be produced from the photosphere through the magnetic dissipation \citep{Giannios2006,Giannios2007,Giannios2008}. \citet{Koers:2007ww} first considered the neutron effects in this model, which was later used by \citet{Meszaros2011} to interpret the GeV time delay.
In the magnetic-dominated but baryon-loaded model \citep{Koers:2007ww,Beloborodov:2009be,Meszaros2011}, MeV photons can escape the plasma at the photosphere radius, which correspond to the  prompt emission. However, GeV photons are produced by the nuclear inelastic collisions between  protons and neutrons at a larger radius. In such two-zone scenario, the strong constraint on the bulk Lorentz factor can be loosened \citep{Hascoet:2011et,Zhao:2010kd,Zou:2010xg}. As pointed out by  \citet{Zhao:2010kd}, the optical depth depends not only on the energy but also on the emission angle. An average Lorentz factor $\Gamma_{\rm min}\approx 600$ can be estimated for GRB 080916c, GRB 090510 and GRB 090902b in the two-zone model. In a similar way, \citet{Zou:2010xg} assumed that the GeV photons were emitted at a larger radius than the MeV photons, and gave an analytical formula for $\Gamma_{\rm min}$ by calculating the optical depth of a GeV photon going through the MeV photons shell. %Since the large optical depth of a GeV photon passing through the background MeV photons can be avoided in the two-zone model, the bulk lorentz factor can be lowered pronouncedly.

In this paper, we use the magnetic-dominated jet model discussed by \citet{Koers:2007ww} and \citet{Meszaros2011} to constrain the bulk Lorentz factor of GRB outflows. We show that the Lorentz factor of the short burst GRB 090510 can be as small as 385, while that of the three long bursts converge to about 260. The rest of the paper is organized as follows: In section \ref{sec:mdjm}, we illustrate the magnetic-dominated jet model and the producing mechanism of MeV and GeV photons briefly. In section \ref{sec:application}, we use the delayed  arrival of GeV photons in  four {\it Fermi} bursts to calculate the bulk Lorentz factor of GRB outflow. In section \ref{sec:discussion}, we discuss the validity of this model. Finally,  conclusions are given in section \ref{conclusion}.

\section{The magnetic-dominated jet model \label{sec:mdjm}}

The hydrodynamics of the GRB outflow depends strongly on its geometry structure. In the  magnetic-dominated jet model,  the Lorentz factor of the outflow increases with radius roughly as \citep{Drenkhahn2002,Drenkhahn2002b,Metzger:2010pp,Granot2011}\footnote{The bulk Lorentz factor of the flow  in the magnetic reconnection model originally took the form $\Gamma(r)\approx \eta (r/r_{\rm sat})^{1/3}$ for $r<r_{\rm sat}$ and $\Gamma(r)\approx \eta$ for $r>r_{\rm sat}$ \citep{Drenkhahn2002,Drenkhahn2002b}, where they considered that the flow starts with the Alfv\'en speed at  the initial radius $r_0$. A compact form $\Gamma(r) \cong (r/r_0)^{1/3}$  was taken by \citet{Koers:2007ww}, where $r_0$ is a length scale defined by specific combination of the parameters. However, a Poynting jet can also be accelerated efficiently without reconnection process \citep{Granot2011}, where $\Gamma$ also takes the form $ \Gamma \sim \sigma_0^{1/3} (r/r_0)^{1/3}$, but $r_0$ denotes the width of the magnetic shell.  \citet{Bosnjak:2011pt} assumed that the format in Eq.(\ref{eq:gamma}) is valid at least in the interval of the Thomson- and pair-production-photosphere radii, and $r_0 $  is roughly the same order of the radius where the jet is launched. Since it was unphysical for the jet to be accelerated to a large speed instantaneously, the Lorentz factor at the base $r_0$ was taken to be of order unity. In the present work, we take the idea of Bo\v{s}njak and Kumar, and write  $\Gamma$ as in Eq.(\ref{eq:gamma}). }
\beq\label{eq:gamma}
\Gamma(r) \simeq \left\{ \begin{array}{ll}
 (r/r_0)^{1/3}~~ &   r \leq r_{\rm sat}, \\
\eta ~~& r > r_{\rm sat},
  \end{array} \right. \eeq
where $r_0$ is assumed to be the base of the outflow, and $r_{\rm sat}$ is the saturation radius. $\eta$ denotes the ratio of the magnetic energy density to the baryon rest mass energy density at $r_0$ initially.

The injected baryons include both protons and neutrons. Initially, the neutron-proton jet accelerates as a single fluid where neutrons and protons have elastic collisions. When the $n-p$ collision time scale is longer than the expansion time scale, the neutron component will coast with a terminal bulk Lorentz factor $\Gamma_n$  at a characteristic radius, while the proton component is still accelerated.  Thus,  the neutron component is embedded in a faster proton flow, and the jet becomes a compound flow \citep{Beloborodov:2009be}.

The cross section of $n-p$ collision  is $\sigma_{\rm nuc} \approx \sigma_{\pi} (c/ v_{\rm rel})$, where $\sigma_{\pi}\approx3\times10^{-26}$  cm$^2$, and $v_{\rm rel}$ is the relative speed of  $p$ to  $n$. When  $v_{\rm rel}\rightarrow c$, the collision is inelastic. This occurs  when the comoving expansion time $t_{\rm exp}'\approx r/2c\Gamma$ becomes shorter than the comoving collision time $t_{\rm nuc}'\approx1/n_p'\sigma_{\pi}c$. Here $n_p'= L x/4\pi r^2 m_p c^3 \eta \Gamma$  is the comoving proton number density, $L$ is the isotropic equivalent luminosity, and $x=n_p/(n_p+n_n)$ is the proton fraction of the baryon density. This gives the characteristic radius $r_{\pi}/r_0 =\eta_{\pi}^6 x/2\eta \Gamma^2$, where $\eta_{\pi}\equiv \left(L \sigma_{\pi}/ \ 4 \pi m_pc^3 r_0 \right)^{1/6} \simeq 1.32\times 10^2 L_{54}^{1/6} r_{0,7}^{-1/6}$. Here we have adopted the $Q=Q_{n} \times 10^n$ convention. Making use of Eq. (\ref{eq:gamma}), one obtains
\begin{eqnarray}
\frac{r_{\pi}}{r_0}= \begin{cases}
  \eta_{\pi}^3 (x\eta_{\pi}/2\eta)^{3/5}~~ &   r < r_{\rm sat}, \\
\eta_{\pi}^6 x/2\eta^3 ~~& r > r_{\rm sat}.
  \end{cases}
\end{eqnarray}

The pion production by the inelastic collisions is inevitable. A certain fraction of energy is carried away by neutrinos, which is an important prediction of the baryon loaded jet model. The $\pi^0$ decay gives primary injected GeV photons. However, these photons undergo $e^{\pm}$ cascades and can not escape the opaque jet. Interactions in the plasma are complex, more details can be found in \citet{Beloborodov:2009be}.

Suppose the final components in the jet contain photons with a Band-like spectrum, and the peak energy is around MeV. These photons start to be emitted when $\tau_T = n'_p \sigma_T r/2\Gamma\sim 1$, which gives the Thomson photosphere radius, i.e., $r_{\rm ph}/r_0 =\eta_{T}^6 /2 \eta \Gamma^2$, where $\sigma_T\approx6.65\times10^{-25}$ ${\rm cm}^2$ is the Thomson cross-section, and $ \eta_{T}\equiv\left(L \sigma_{T} /4 \pi m_pc^3 r_0 \right)^{1/6}\simeq 2.22\times 10^2 L_{54}^{1/6} r_{0,7}^{-1/6}$.
Using Eq. (\ref{eq:gamma}) for $\Gamma$, one obtains
\beq
\frac{r_{\rm ph}}{r_0} = \left\{ \begin{array}{ll}
 \eta_T^3 (\eta_T /2 \eta)^{3/5}~~ &   r < r_{\rm sat}, \\
 \eta_T^6 /2 \eta^{3} ~~&   r > r_{\rm sat}. \\
  \end{array} \right.
\eeq

The simulation of magnetohydrodynamics (MHD) shows that the jet can form a conical structure \citep{Tchekhovskoy:2008}. After the jet exits the stellar envelope, the inner jet runs faster than the outer sheath. Thus, the Lorentz factor tapers off towards the edges. In such structure, the neutrons from the outer sheath can drift into the inner core. The relative radial Lorentz factor ratio between neutrons and baryons is larger than 1, which ensures that the collisions are inelastic \citep{Meszaros2011}.  Suppose the jet has an open angle $\theta$, the transverse pion optical depth can be expressed as
$\tau_{\pi, \bot} \approx n_p' \sigma_{\pi} r \theta\Gamma=\eta_{\pi}^6(r_0/r)(x\theta/\eta).$
The jet becomes transversely optically thin ($\tau_{\pi, \bot}=1$) at $r_{\pi, \bot}$, which is defined as
\beq \frac{r_{\pi, \bot}}{r_0} = \eta_{\pi}^6 \frac{x\theta}{\eta}. \eeq

The dynamical evolution of the jet depends on $\eta$. If $\eta$ is large, for instance $\eta=600\eta_{600}$, the saturation radius $r_{\rm sat}$ may be larger than $r_{\pi}$, $r_{\rm ph}$ and $r_{\pi, \bot}$. Making use of $\Gamma=(r/r_0)^{1/3}$, one obtains the following characteristic radii:
\begin{eqnarray}\label{radii1}
\begin{cases}
r_{\pi} \simeq  4.04 \times 10^{12}   L_{54}^{3/5} r_{0,7}^{2/5}x_{0.5}^{3/5} \eta_{600}^{-3/5}~~{\rm cm}, \\
r_{\rm ph} \simeq 3.98 \times 10^{13}  L_{54}^{3/5} r_{0,7}^{2/5} \eta_{600}^{-3/5}~~{\rm cm}, \\
r_{\pi, \bot} \simeq 4.41 \times 10^{14}  L_{54} \eta_{600}^{-1}x_{0.5} \theta_{-2}~~{\rm cm}, \\
r_{\rm sat} \simeq  2.16\times 10^{15} r_{0,7} \eta_{600}^3~~{\rm cm}. \\
\end{cases}
\end{eqnarray}
On the other hand, if $\eta$ is small, $r_{\rm sat}$ may become smaller than $r_{\pi}, r_{\rm ph}$ and $r_{\pi, \bot}$. In this case, one obtains the corresponding radii:
\begin{eqnarray}\label{radii2}
\begin{cases}
r_{\pi}  \simeq  1.32 \times 10^{13}   L_{54}x_{0.5} \eta_{100}^3~~{\rm cm}, \\
r_{\rm ph} \simeq 5.98 \times 10^{14}  L_{54}\eta_{100}^3~~{\rm cm}, \\
r_{\pi, \bot}  \simeq 2.64 \times 10^{15}  L_{54} \eta_{100}^{-1}x_{0.5} \theta_{-2}~~{\rm cm}, \\
r_{\rm sat}  \simeq  1.00\times 10^{13} r_{0,7} \eta_{100}^3 ~~{\rm cm}. \\
\end{cases}
\end{eqnarray}
Here  $\eta$  is in unit of 100, i.e., $\eta=100\eta_{100}$. In both cases above, one has the order $r_{\pi}< r_{\rm ph}<r_{\pi, \bot}$.

The GeV photons are assumed to be produced at $r_{\pi, \bot}$ by transverse nuclear collisions. However, the large optical depth prevents them to escape immediately. The spectra of produced photons depend on many parameters. Suppose that the Band spectrum peaks at $E_p\sim 1$ MeV, then the optical depth seen by a photon with energy $E$ at radius $r$ is approximately given by \citep{Beloborodov:2009be}
\begin{equation} \label{eq:tauEr}
  \tau_{\gamma\gamma}(E,r)\approx\frac{2\times 10^5}{40^{-\beta-1}}r_{12}^{-1}L_{54}\bigg{(}\frac{E}{10~{\rm GeV}}\bigg{)}^{-\beta-1}\eta_{600}^{2\beta},
\end{equation}
where $\beta\approx-2.5$ is the spectrum index above the peak energy $E_p$, and $r_{12}=r/10^{12}$ cm.
Setting $\tau_{\gamma\gamma}=1$, we get the $\gamma-\gamma$ transparency radius,
\begin{equation}\label{rgg1}
r_{\gamma\gamma}(E) \approx 2.50\times 10^{13} E^{3/2} \eta_{600}^{-5} L_{54} ~~ {\rm cm}.
\end{equation}
Or equivalently, $r_{\gamma\gamma}(E) \approx 1.94\times 10^{17} E^{3/2} \eta_{100}^{-5} L_{54} $ cm for the  $\eta \sim 100$ case.
Here, $E$ is the observed photon energy in unit of GeV. Thus, at $r_{\pi, \bot}$, multi-GeV photons will be copiously produced by the transverse indrift neutrons colliding with jet core baryons. In the radius range $r_{\pi, \perp} < r < r_{\gamma \gamma}$, these photons will annihilate into electron-positron pairs due to the large optical depth. Only beyond the radius $r_{\gamma \gamma}$, the produced GeV photons can escape without obstructions.

The time delay for a photon with energy $E$ relative to the onset time of MeV photons, equates to the time it takes for the jet to propagate from $r_{\rm ph}$ to $r_{\gamma\gamma}(E)$,
\begin{eqnarray}
  \Delta t=(1+z)\int_{r_{\rm ph}}^{r_{\gamma\gamma}(E)}\frac{dr}{2\Gamma^2c},
\end{eqnarray}
where $z$ is the redshift. The explicit formula for $\Delta t$ depends on the order of $r_{\rm ph}$, $r_{\rm sat}$ and $r_{\gamma\gamma}$. From Eqs. (\ref{radii1}), (\ref{radii2}) and (\ref{rgg1}), one can obtain $r_{\rm ph}\lesssim r_{\gamma\gamma}$  for the generic $E\gtrsim 10$ GeV and $\eta\gtrsim 100$.
Using Eq. (\ref{eq:gamma}) for $\Gamma$, one obtains the formats of $\Delta t$ for three different cases:
\begin{eqnarray}\label{delta_t}
  \Delta t\approx\begin{cases}
  \frac{3(1+z)r_0}{2c}\big{[}(\frac{r_{\gamma\gamma}}{r_0})^{\frac{1}{3}}-(\frac{r_{\rm ph}}{r_0})^{\frac{1}{3}}\big{]} &  r_{\rm ph}< r_{\gamma\gamma}<r_{\rm sat};\\
    \frac{r_{\gamma\gamma}(E)-r_{\rm ph}}{2\eta^2c}(1+z) &    r_{\rm sat}<r_{\rm ph}<r_{\gamma\gamma};\\
  \frac{3(1+z)r_0}{2c}\big{[}(\frac{r_{\rm sat}}{r_0})^{\frac{1}{3}}-(\frac{r_{\rm ph}}{r_0})^{\frac{1}{3}}\big{]} & \phantom{r_{\rm sat}<r_{\rm ph}<r_{\gamma\gamma}} \\
  \phantom{00}+\frac{r_{\gamma\gamma}(E)-r_{\rm sat}}{2\eta^2c}(1+z)  & \ r_{\rm ph}<r_{\rm sat}<r_{\gamma\gamma}.\\
  \end{cases}
\end{eqnarray}

\begin{figure}
\centering
  \plotone{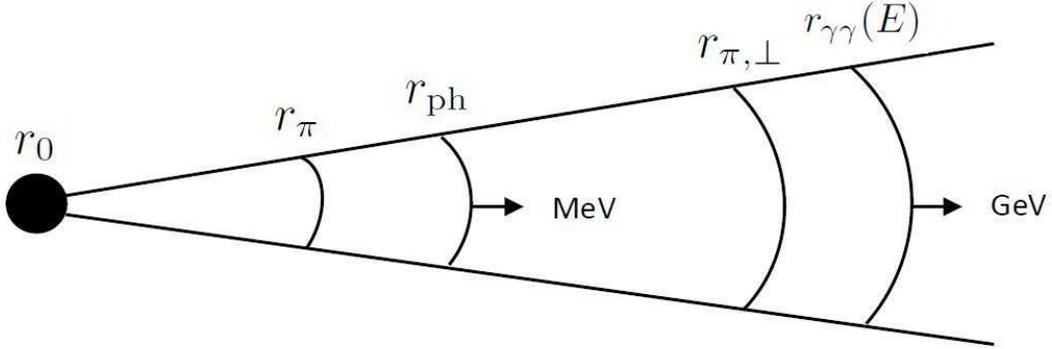}
  \caption{A phenomenological description of the magnetic-dominated jet model. $r_{\rm sat}$ can locate before $r_{\rm ph}$, or beyond $r_{\gamma\gamma}$, or at any place between them.} \label{fig:jetmodel}
\end{figure}

A phenomenological illustration of the magnetic-dominated jet  model is depicted in Fig. \ref{fig:jetmodel}. A  magnetic-dominated but baryon-loaded jet is launched from a progenitor at the initial radius $r_0$. The bulk Lorentz factor of the jet evolves as  Eq.(\ref{eq:gamma}). MeV photons are produced at the radius $r_{\pi}$ by nuclear collisions, $\pi_0$-decay, electron-positron annihilation, magnetic dissipation, and synchrotron radiation, etc.. But these photons can only escape at the photosphere radius $r_{\rm ph}$, where the Thomson optical depth decreases to below unity. Thus, multi-MeV photons are emitted at $r_{\rm ph}$ and lead to the observed Band-type spectra \citep{Veres:2012sb}. The GeV photons are assumed to produce at $r_{\pi,\bot}$ by transverse drift nuclear collisions, inverse Compton radiation, etc.. But these GeV photons are capable to escape only at a larger radius $r_{\gamma\gamma}$ due to the large optical depth at $r_{\pi,\bot}$. The time it takes for the jet to propagate  from $r_{\rm ph}$ to $r_{\gamma\gamma}$ naturally leads to the GeV time delay relative to the onset of MeV photons.

\section{Constraints on the Lorentz factors \label{sec:application}}

From Eqs. (\ref{radii1}), (\ref{radii2}), (\ref{rgg1}) and (\ref{delta_t}), the time delay $\Delta t$ strongly depends on the terminal bulk Lorentz factor $\eta$.
%Taking the $r_{\rm ph}< r_{\gamma\gamma}<r_{\rm sat}$ case for example, one has $\Delta t \approx (1+z)(0.068E^{1/2}L_{54}^{1/3}\eta_{600}^{-5/3}
%r_{0,7}^{2/3}-0.079L_{54}^{1/5}\eta_{600}^{-1/5} r_{0,7}^{4/5})$.
Inspired by this fact, we make use of the magnetic-dominated jet model discussed above to calculate $\eta$ for four {\it Fermi} bursts, GRB 080916c, GRB 090510, GRB 090902b and GRB 090926, respectively.

The observed parameters which are necessary in the calculation are listed in Table \ref{tab:observedvalue}. Note that GRB 090510 is a short burst, while the other three are long bursts. In Table \ref{tab:observedvalue}, $E_{\rm high}$ was taken to be the energy of the most energetic photon in each GRB. One exception is that the second energetic photon with $E_{\rm high}=11.16$ GeV in GRB 090902b was chosen, while the most energetic $33.4$ GeV photon arriving at $82$ s was excluded. This is because the isolated photon is far apart from other GeV photons and it is quite possible that this individual event happened when the jet encountered the interstellar medium.

\begin{deluxetable}{cccccc}
\tablecaption{\label{tab:observedvalue}}
%\tabletypesize{\scriptsize}
\tablewidth{0pt}
\tablehead{\colhead{GRB}&\colhead{E$_{\rm ios,54}$}&\colhead{T$_{90}$}&\colhead{z}&\colhead{$E_{\rm high}$}&\colhead{$\Delta t_{\rm obs}$}}
\startdata
080916c~~&8.8~~&66~~&4.35 ~~&13.22~~&12.94 \\
090510  ~~&0.11~~ & 0.6~~ &0.90 ~~ &31.0~~& 0.20\\
090902b ~~ &3.7 ~~ &22 ~~ &1.82 ~~ &11.16~~ &9.5\\
090926  ~~&2.2~~ &13 ~~&2.11 ~~ &19.6 ~~&21.5\\
\enddata
\tablecomments{The observed parameters of four $Fermi$-detected GRBs. $E_{\rm ios,54}$ is the isotropic equivalent energy in unit of $10^{54}$ ergs. T$_{90}$ is $90$\% the GRB duration time in unit of second. $z$ is the GRB redshift. $E_{\rm high}$ is the highest energy of photons for each burst in unit of GeV. $\Delta t_{\rm obs}$ is the observed time delay between the highest energy photon relative to the onset of 100 MeV photons. The data were taken from \citet{Chang:2012gq}.}
\end{deluxetable}

There are several parameters which are uncertain, such as the initial radius $r_0$, the jet open angle $\theta$ and the ratio of proton number density to that of  baryons $x$. Long bursts usually have time variability $\delta t\approx 10$ ms, thus the initial radius is taken to be $r_0\approx c\delta t\approx10^8$ cm for long bursts. The value of $r_0$ for short bursts is usually assumed to be smaller than that of the long bursts, and we set $r_0\approx 10^7$ cm for short burst GRB 090510. A nominal value of jet open angle $ \theta$ is taken to be  0.01, and the proton fraction of the baryon density $x$ is approximately 0.5, i.e., $\theta_{-2}\approx 1$ and $x_{0.5}\approx 1$ \citep{Meszaros2011}. When $x$ approaches zero, this model reduces to the magnetic jet model without the loaded baryons \citep{Bosnjak:2011pt}.

As was mentioned in  section \ref{sec:mdjm}, the explicit formula for $\Delta t$ depends on the order of $r_{\rm ph}$, $r_{\rm sat}$ and $r_{\gamma\gamma}$, which is not known previously. Thus, a self-consistent calculation should be taken carefully.

{\bf Case I}: First, we consider the $r_{\rm sat}>r_{\gamma\gamma}>r_{\rm ph}$ case (see the first formula in Eq.(\ref{delta_t})). The calculated saturation bulk Lorentz factor and characteristic radii are listed in Table \ref{tab:r<rsat}. The characteristic radii of the short burst GRB 090510 follow the order $r_{\rm sat}>r_{\gamma\gamma} > r_{\rm ph}$, which is self-consistent. However, for three long bursts, the results indicate $r_{\rm sat}<r_{\gamma\gamma}$, which are in contradiction with the assumption. The short burst GRB 090510 has a Lorentz factor about $720$, which is much lower than the prediction of one-zone models. For instance,  \citet{Abdo2009b} presented that the bulk Lorentz factor of GRB 090510 was as large as $1200$. Our results indicate that GeV photons in the short burst are emitted before the Lorentz factor of the jet saturates.

\begin{deluxetable}{ccccccc}
\tablecaption{ \label{tab:r<rsat}}
%\tabletypesize{\scriptsize}
\tablewidth{0pt}
\tablehead{\colhead{GRB}&\colhead{$\eta_{600}$}&\colhead{$r_{\rm ph}$/cm}&\colhead{ $r_{\gamma\gamma}$/ cm}&\colhead{ $r_{\rm sat}$/cm}&\colhead{$r_{\pi}$/cm}&\colhead{$r_{\pi,\bot}$/cm}}
\startdata
080916c~~ &0.39~~ & $5.24\times 10^{13}$~~ & $1.78\times10^{16}$ ~~& $1.28\times10^{15}$ ~~& $5.33 \times 10^{12}$~~& $1.51\times 10^{14}$\\
090510 ~~&1.20 ~~& $1.29\times 10^{13}$~~ & $3.18\times10^{14}$~~ &  $3.73\times10^{15}$ ~~&$1.31 \times 10^{12}$~~& $6.74 \times 10^{13}$\\
090902b ~~&0.32~~ & $6.80\times 10^{13}$ ~~ & $4.67\times10^{16}$~~ &$7.08\times10^{14}$ ~~&$6.90\times 10^{12}$~~& $2.32\times 10^{14}$ \\
090926~~ &0.26~~ & $7.73\times10^{13}$ ~~& $3.09\times10^{17}$~~ & $3.80\times10^{14}$ ~~&$7.84\times 10^{12}$~~& $ 2.87 \times 10^{14}$ \\
\enddata
%\tablenotemark{Note}
\tablecomments{The calculated saturation bulk Lorentz factors and characteristic radii under the assumption that $r_{\rm sat}>r_{\gamma\gamma}>r_{\rm ph}$. We choose $r_{0,7}=1$ for short burst GRB 090510 and $r_{0,7}=10$ for other three long bursts. The characteristic radii of three long bursts are not self-consistent.}
\end{deluxetable}

{\bf Case II}: Then, we consider the $r_{\rm sat}<r_{\rm ph}<r_{\gamma\gamma}$ case (see the second formula in Eq.(\ref{delta_t})).  The results are given in Table \ref{tab:r>rsat}. For all of the four bursts, we have $r_{\pi,\bot}<r_{\rm ph}$. Thus, the transverse nuclear collisions happen inside the photosphere, and GeV photons are converted to the $e^{\pm}$  cascades. The overlap of the producing regimes of  GeV and  MeV photons is possible, but GeV photons are  attenuated until $r_{\gamma \gamma}$. The spectrum of GRB 090510 is fitted well by the Band function plus a power-law component which dominates in the band above 30 MeV \citep{Zhao:2010kd}, this can be explained well by the Magnetic-dominated jet model. If these arguments are true, the bulk Lorentz factor of GRB 090510 is further reduced to $385$. In Table \ref{tab:r>rsat}, one also notice  that for GRB 080916c, $r_{\rm ph}<r_{\rm sat}$, which is not self-consistent. The bulk Lorentz factor of GRB 090902b and GRB 090926 in this case  are calculated to be 245 and 252, respectively.

\begin{deluxetable}{ccccccc}
\tablecaption{\label{tab:r>rsat}}
%\tabletypesize{\scriptsize}
\tablewidth{0pt}
\tablehead{\colhead{GRB}&\colhead{$\eta_{100}$}&\colhead{$r_{\rm ph}$/cm}&\colhead{ $r_{\gamma\gamma}$/ cm}&\colhead{ $r_{\rm sat}$/cm}&\colhead{$r_{\pi}$/cm}&\colhead{$r_{\pi,\bot}$/cm}}
\startdata
080916c~~ &2.58~~ & $1.37\times10^{15}$ ~~& $1.09\times10^{16}$~~ &$1.72\times10^{15}$~~&$5.02 \times 10^{12}$~~& $1.36\times 10^{14}$\\
 090510 ~~&3.86 ~~& $6.31\times10^{15}$ ~~& $7.16\times10^{15}$~~ &$5.75\times10^{14}$ ~~&$1.91 \times 10^{12}$ ~~& $1.25 \times 10^{14}$\\
090902b~~ &2.45 ~~& $1.48\times10^{15}$ ~~ & $1.38\times10^{16}$~~ &$1.47\times10^{15}$~~&$5.96\times 10^{12}$~~& $1.81 \times 10^{14}$\\
090926~~ &2.52~~ & $1.62\times10^{15}$~~ & $2.80\times10^{16}$ ~~& $1.60\times10^{15}$~~&$5.88\times 10^{12}$~~& $1.77\times 10^{14}$\\
\enddata
%\tablenotemark{Note}
\tablecomments{The calculated saturation bulk Lorentz factors and characteristic radii under the assumption that $r_{\rm sat}<r_{\rm ph}<r_{\gamma\gamma}$. The parameters are the same as in Table \ref{tab:r<rsat}. The characteristic radii of the GRB 080916c are not self-consistent.}
\end{deluxetable}

 {\bf Case III}: Finally, we consider the $r_{\rm ph}<r_{\rm sat}<r_{\gamma\gamma}$ case (see the third formula in Eq.(\ref{delta_t})). The results are listed in Table \ref{tab:rph<rsat<rgg}. The data of GRB 090510 are absent, because any value of $\eta$ can not fit $\Delta t_{\rm obs}=0.2$ s by the formula. The minimal value of $\Delta t$ is $0.34$ s locating at $\eta_{600}\approx 1$. The data of three long bursts fit well in this case. The bulk Lorentz factors are 270, 252 and 258 for GRB 080916c, GRB 090902b and GRB 090926, respectively.

\begin{deluxetable}{ccccccc}
\tablecaption{\label{tab:rph<rsat<rgg}}
%\tabletypesize{\scriptsize}
\tablewidth{0pt}
\tablehead{\colhead{GRB}&\colhead{$\eta_{600}$}&\colhead{$r_{\rm ph}$/cm}&\colhead{ $r_{\gamma\gamma}$/ cm}&\colhead{ $r_{\rm sat}$/cm}&\colhead{$r_{\pi}$/cm}&\colhead{$r_{\pi,\bot}$/cm}}
\startdata
080916c~~ &0.45~~ & $4.82\times10^{13}$~~ & $8.68\times10^{15}$~~ & $1.97\times10^{15}$~~& $4.89 \times 10^{12}$ ~~&$1.31 \times 10^{14}$ \\
090902b~~ &0.42~~ & $5.77\times10^{13}$~~ & $1.20\times10^{16}$~~ & $1.60 \times10^{15}$~~& $5.86\times 10^{12}$ ~~&$1.77 \times 10^{14}$ \\
090926~~ &0.43~~ & $5.71\times10^{13}$~~ & $2.50 \times10^{16}$~~ & $1.72\times10^{15}$~~& $5.80 \times 10^{12}$ ~~&$1.74 \times 10^{14}$ \\
\enddata
%\tablenotemark{Note}
\tablecomments{The calculated saturation bulk Lorentz factor and characteristic radii for three long bursts under the assumption that $r_{\rm ph}<r_{\rm sat}<r_{\gamma\gamma}$. The parameters are the same as in Table \ref{tab:r<rsat}.}
\end{deluxetable}

\begin{deluxetable}{ccccc}
\tablecaption{\label{tab:LorentzFactor}}
 \tablewidth{0pt}
\tablehead{\colhead{\phn}&\colhead{080916c}&\colhead{090510}&\colhead{090902b}&\colhead{ 090926}}
  \startdata
 \makebox[3.5cm-1\tabcolsep][c]{$r_{\rm ph}<r_{\gamma\gamma}<r_{\rm sat}$}&{$\times$}&{720}&{$\times$}&{$\times$}\\ %\hline
 \makebox[3.5cm-1\tabcolsep][c]{$r_{\rm sat}<r_{\rm ph}<r_{\gamma\gamma}$}&{$\times$}&{385}&{245}&{252}\\ %\hline
 \makebox[3.5cm-1\tabcolsep][c]{$r_{\rm ph}<r_{\rm sat}<r_{\gamma\gamma}$}&{270}&{$\times$}&{252}&{258}\\ \hline
 \enddata
\tablecomments{The saturation bulk Lorentz factors for the four {\it Fermi}-detected bursts in three different cases. ``$\times$'' denotes the inconsistent case.}
\end{deluxetable}

So far, we have obtained the self-consistent Lorentz factors and characteristic radii for all of the four bursts, and summarized the Lorentz factors in Table \ref{tab:LorentzFactor}.
From the calculation above, the two conditions $r_{\pi} < r_{\rm ph}$ and $r_{\pi,\bot}<r_{\gamma \gamma}$ always hold.
For short burst GRB 090510, the self-consistent cases are case I and II, and the corresponding Lorentz factors are 720 and 385, respectively. For the three long bursts, case III is valid for all of them, but they are all excluded in the case I. The allowed Lorentz factors for the three long bursts seem to converge to an average value about $260$. The saturation bulk Lorentz factor of the short burst GRB 090510 is reduced sharply, but still higher than that of long bursts. Another interesting feature is that GeV photons are emitted after the bulk Lorentz factor saturates for long bursts, i.e. in the coasting phase. For the short burst, however, both MeV and GeV photons are emitted either in the expansion phase or in the coasting phase. Case I can not happen for the long bursts, while case III can not happen for the short bursts.

\section{Discussion \label{sec:discussion}}

%%%%%%%%%%%%%%%%%%%%%%%%%%%%%%%%%%%%%%%%%%%%%%%%%%%%%%%%%%%%%%%%%%%%%%%%%%%%%%%%%%%%%%%%%%%%%%
Besides the time delay, another important feature of GeV emissions is that they last much longer than the sub-MeV photons \citep{Gao:2009,Kumar:2009b,Ghirlanda:2010b,Ghisellini2010}. For instance, the duration time of the sub-MeV photons is 55 seconds in GRB 080916c, while photons with energy $>100$ MeV last about 1400 seconds \citep{Kumar:2009b}. The observed decline of flux can be explained by the synchrotron radiation in the external shock (ES), i.e., $F_{\nu} \propto t^{(3 \beta +2)/4}\nu^{\beta/2}$ ($\beta=-2.4$ for GRB 080916c). The data of the initial 55 seconds are able to explain the observed X-ray and optical flux of the afterglow one day later. Thus, the GeV emissions have an afterglow origin.

The spectrum and the light curve of the GRB 090510 were also explained by the synchrotron radiation in the ES model \citep{Ghirlanda:2010b}. \citet{Ghisellini2010} studied the light curves of 11  GRBs detected by LAT, and concluded that LAT fluxes decay in a common way $F_{\nu} \propto t^{-1.5}$ for the four brightest GRBs studied in this paper. The LAT fluxes can be interpreted as the fireball emission in the radioactive regime. The spectra of the GeV emissions in some bursts showed a different power law from the Band function. Thus, the spectra and the light curves present strong evidences that GeV emissions have different origin with the sub-MeV emissions.
As hinted by \citet{Ghisellini2010}, one can divide the ``total emission time" of sub-MeV and GeV emissions into two parts\footnote{X-ray and optical radiation usually arrive one day later, i.e., in the afterglow phase. We do not include them in the ``total emission time".}: one is the overlap regime where both the sub-MeV  and GeV photons are present; another is the regime where only LAT photons exist. The latter can be named as the early afterglow.

Notice that our calculation about GeV time delay is valid in the overlap regime, the GeV emissions in the early afterglow are not discussed. Once the outflow collides with the environment medium, the forward shock can also occur in the magnetic-dominated jet model. Both electrons and protons can be accelerated by the shock and form a power law spectral distribution. The characteristic frequency of the synchrotron radiation follows $\nu = \Gamma \gamma_{e,p}^2 q B/2 \pi m_{e,p} c$, the synchrotron radiation of protons can be ignored compared to that of electrons. The produced photons have the same power law spectrum with electrons. Since the optical depth is small at so large radius, photons are emitted immediately. The radiative fireball leads to the long duration time and light curve of  GeV photons. Therefore, synchrotron radiation in the ES can explain the GeV emissions of the early afterglow in the magnetic-dominated jet model.

Now we consider the spectra and light curve in the overlap regime. In the neutron rich environment, the inelastic collisions produce pions, which further decay into photons with minimal energy $70 \Gamma $ MeV \citep{Fan:2008}. In the meantime, the produced neutrinos can escape with the observed energy $\sim 0.1 \Gamma$ GeV \citep{Beloborodov:2009be}. However, these high energy neutrinos are difficult to be detected on the earth. \citet{Koers:2007ww} estimated that less than 1 GRB neutrino event can be detected every year for nominal GRB parameters ($z=1$). Thus, one can not exclude the baryon loaded model by the neutrino argument. If the spectrum of the protons is in power law, the resulted photons will also follow the same distribution.  However, these original photons will quickly convert to $e^{\pm}$ via $\gamma \gamma$ reaction unless they are produced at a large radius where optical depth is below unity. The subsequent processes  including Coulomb and Compton interactions are complex. Finally a Band like spectrum can form and the radiation becomes the observed  photons in prompt phase \citep{Beloborodov:2009be}.

Without the neutron component, the magnetic dominated outflow can  dissipate energy efficiently \citep{Drenkhahn2002,Drenkhahn2002b}. A non-thermal spectrum can be produced by the magnetic reconnection model, and this spectrum is close to the observed prompt GRB emission \citep{Giannios2006,Giannios2007}. This means a broken power law (Band-like) spectrum can be produced with or without baryons. The luminosity of GeV emissions in the overlap regime rose as $L \propto t^2$ in most GRBs \citep{Ghisellini2010}. One exception is GRB 080916c, where the luminosity rose as $L \propto t^6$, which was a puzzle \citep{Kumar:2009b}. Our conclusion that the long bursts have the order $r_{\rm ph}< r_{\rm sat}<r_{\gamma \gamma}$ may help to understand this puzzle. In this order, the jet is still in the expansion phase after the prompt emission, which means that $\Gamma$ increases with time. If the jet is magnetic-dominated, one has $\Gamma \sim (r/r_0)^{\mu}$ and $r=2a c t \Gamma^2$. In this way, $\mu=1/4$ leads to $\Gamma \propto t^{1/2}$. Since $L \propto t^2 \Gamma^8$ \citep{Ghisellini2010}, one can explain the puzzle. The light curve of the GeV emissions in the overlap regime strongly favors the magnetic jet model. \citet{Gao:2009} also found that the physical composition of the GRB 080916c is likely magnetic.

The spectra of GeV photons do not evolve with time, and have a flatter component (the slope intermediate between $\alpha$ and $\beta$ of the Band function) \citep{Ghisellini2010}. These evidences strongly indicate that the GeV photons have a different producing mechanism. The two component GRB spectra were discussed by \citet{Veres:2012sb} recently, where a dissipative photosphere gives the prompt MeV emission, while GeV emission are produced by the inverse Compton scattering. The model studied in this work also belongs to the two component case, many possible spectra are able to account for different  GRBs. Therefore, the magnetic dominated jet model can explain many phenomenon of GRBs, such as the GeV time delay, the light curves and the spectra, etc..

\section{Conclusion \label{conclusion}}

In this paper, we  have studied the  bulk Lorentz factor of GRB outflow within the framework of magnetic-dominated jet model. We found that the emission mechanisms of the short and long GRBs are different. The long bursts have a unified bulk Lorentz factor around $260$ in both case II and III. However, the Lorentz factor of the short burst is 720 in case I and 385 in case II. These values are much smaller than that obtained from the ``one-zone" scenario. \citet{Zhao:2010kd} calculated the Lorentz factor of GRB 080916c, GRB 090510 and GRB 090902b, and showed that $\eta \sim 600$ could be consistent with observations in the ``two-zone" scenario. Their values were still $2-3$ times larger than our result for long bursts. They also proposed that the Lorentz factor could be even lowered in the ``multi-zone" scenario. The magnetic-dominated jet model discussed here is a kind of ``multi-zone" model, according to which photons with higher energy are emitted at a larger radius. The Lorentz factors for long bursts  we obtained here were well inside the limits ($\sim 200-400$) given by \citet{Zou:2010xg}.

\begin{figure}
  \centering
 \plotone{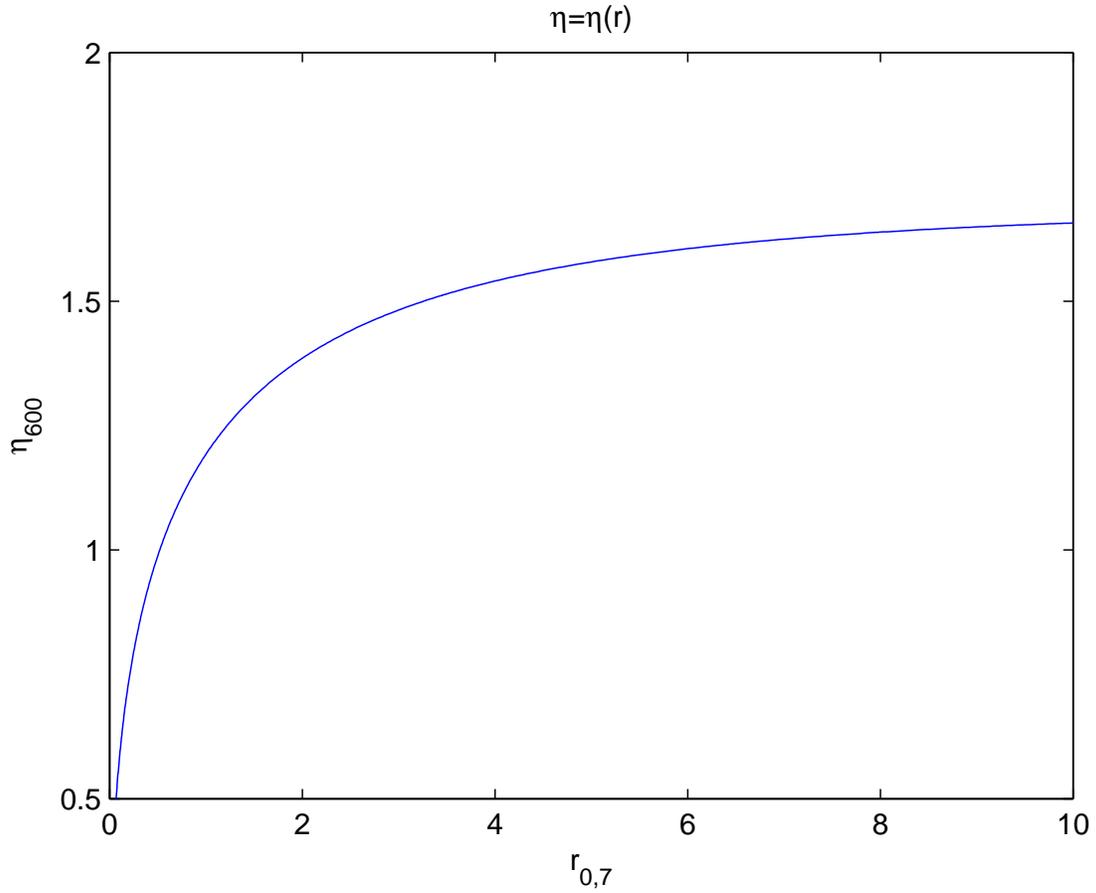}
  \caption{\small{The relation between the saturation bulk Lorentz factor $\eta$ and the initial radius $r_0$ for the short burst GRB 090510.}}
  \label{fig:eta-r}
\end{figure}

According to the magnetic-dominated jet model, the Lorentz factor depends on the initial radius $r_0$,  which is an undetectable parameter. In case I, only short burst GRB 090510 is self-consistent. As indicated in Figure \ref{fig:eta-r}, $\eta_{600}$ is asymptotic to 1.7 when $r_{0,7}$ goes to infinity. This means that the maximal Lorentz factor of GRB 090510 is about 1000  in this model.  A small $r_0$ leads to a small $\eta$. For instance, if $r_{0,7}=0.1$, $\eta$ becomes $360$. In case II, $\eta_{600}$ does not depend on $r_{0,7}$, because the formulas of  $r_{\gamma \gamma}$ and $r_{\rm ph}$ are independent of $r_{0,7}$ (see Eq. (\ref{radii2}) and (\ref{rgg1})). In case III, $\eta_{600}$ depends weakly on $r_{0,7}$, since $r_{\gamma \gamma}$ are $2 \sim 3$ orders of magnitude larger than $r_{\rm ph}$. Thus, $r_{\rm ph}$ can be ignored in the calculation. The strong correlation between $\Delta t$ and $\eta$ has an advantage: a small variation of $\Delta t$ will not lead to a big change of $\eta$.

The time delay of GeV photons relative to the MeV photons can be well explained in the magnetic-dominated jet model. The bulk Lorentz factors of both long and short GRBs are reduced significantly. For GRB 090510, the possible minimal Lorentz factor is 385. For three long bursts GRB 080916c, 090902b and 090926, the Lorentz factors converge to about 260. The Lorentz factor of short burst is still larger than that of the long bursts. One common feature of the long bursts is that GeV photons are emitted after the bulk Lorentz factor saturates. In contrast, GeV photons in short burst can be emitted either in the expansion phase or in the coasting phase, and the bulk Lorentz factor in the former case is about one time larger than that in the latter case.

The fact that three long bursts have a common Lorentz factor may imply that the long bursts have the same origin. One prevalent idea is that long GRBs are caused by the collapse of a massive star (such as Wolf-Rayet star) \citep{woosley1993,paczynski1998,woosley2006}. The short duration time and the large Lorentz factor of the short burst may imply a different kind of central engine mechanism. For instance, short GRBs can originate from the merger of two compact objects (such as NS-NS binary system and NS-BH binary system) \citep{Goodman1986,meszaros1992,zhang2006}.

%%%%%%%%%%%%%%%%%%%%%%%%%%%%%%%%%%%%%%%%%%%%%%%%%%%%%%%%%%%%%%%%%%%%%%%%%%%%%%%%%%%%%%%%%%%%%%%
\begin{acknowledgments}
We are grateful to M. H. Li, X. Li and S. Wang for useful discussions. This work has been funded in part by the National Natural Science Fund of China under Grant No. 10875129 and No. 11075166. The work of Y.~G.~Jiang is also funded by  the China Postdoctoral Science Foundation funded project (Grant No. 2012M510548).
\end{acknowledgments}

\end{document}